**Excited-state spectroscopy using single-spin manipulation in diamond**

G. D. Fuchs[1], V. V. Dobrovitski[2], R. Hanson[3], A. Batra[4], C. D. Weis[4], T. Schenkel[4], and D. D. Awschalom[1]

[1]Center for Spintronics and Quantum Computation, University of California, Santa Barbara, California 93106

[2]Ames Laboratory and Iowa State University, Ames, Iowa 50011

[3]Kavli Institute of Nanoscience, Delft University of Technology, P.O. Box 5046, 2600 GA Delft, Netherlands

[4]Lawrence Berkeley National Laboratory, Berkeley, California, 94720

Abstract

We use single-spin resonant spectroscopy to study the spin structure in the orbital excited-state of a diamond nitrogen-vacancy center at room temperature. We find that the excited state spin levels have a zero-field splitting that is approximately half of the value of the ground state levels, a *g*-factor similar to the ground state value, and a hyperfine splitting ~20x larger than in the ground state. In addition, the width of the resonances reflects the electronic lifetime in the excited state. We also show that the spin-splitting can significantly differ between NV centers, likely due to the effects of local strain, which provides a pathway to control over the spin Hamiltonian and may be useful for quantum information processing.

PACS numbers: 76.30.Mi, 0.3.67.Lx, 42.50.Ex

Nitrogen-vacancy (NV) centers in diamond have numerous properties that make them an attractive solid-state system for fundamental investigations of spin-based phenomena as well as for the implementation of quantum information processing. Among these are a long spin coherence time at room temperature [1], an electronic level structure that allows optical initialization as well as read-out [2, 3], and the ability to address individual centers [4]. These properties have been used to study the interactions with nearby electronic [1, 5, 6] and nuclear spins [7, 8, 9]. Despite recent progress in understanding the details of the electronic and spin structure of this defect center [10, 11] there are still important unresolved questions; particularly in regards to the detailed structure of the excited state (ES) [12]. Understanding these excited state spin levels is of critical importance for high-speed optical control of diamond NV centers [13] and for potential quantum information applications such as a quantum repeater [14].

Accessing the spin structure of the orbital excited state has been an experimental challenge due in part to the short lifetime [10, 12]. Furthermore, local strain is expected to play a more important role in the excited state than it does in the ground state [10, 15] which means that ensemble measurements are problematic whenever there is a variation of strain within the sample. Previous experiments designed to probe the excited state have focused on resonant optical transitions at low temperature [15, 16] where the zero-phonon transition line is narrow and the vibronic sidebands can be excluded. Here, we pursue a simple and powerful alternative approach to probe the excited state spin levels at room temperature where we can easily vary the strength and direction of the magnetic field. By applying frequency-domain spin resonance techniques using a large microwave driving field, we can induce spin rotations of a single NV spin during the short (~10 ns) period in the ES, allowing us to capture the ES level structure as a function of applied magnetic field. Using this single-spin resonance spectroscopy, we determine room temperature values of the ES longitudinal splitting coefficient, $D_{es}$=1.425±0.003 GHz, the strain-induced transverse anisotropy constant $E_{es}$=70±30 MHz, the ES g-factor

$g_{es}$=2.01±0.01, and the constant of hyperfine coupling with the spin-1/2 nucleus of [15]N, $A_{es}$=61±6 MHz. Finally, we demonstrate significant differences between the spin levels for different NV centers presumably due to variations in local strain.

The NV center is a defect formed when a substitutional nitrogen is located at a lattice site adjacent to a carbon vacancy. In the negatively charged state (NV-), as studied here, this defect center is a spin triplet in both the electronic ground state (GS, $^3$A) as well as the ES ($^3$E) (Fig. 1.(a)). Optical transitions between these levels are primarily spin-conserving [10] except for the presence of a non-radiative and spin-selective relaxation mechanism through an intersystem crossing to $^1$A that allows for initialization under optical pumping into the $m_S$=0 spin state [10, 17]. The spin-selective nature of this process also creates a difference in the photoluminescence signal that depends on the spin state. In particular, the photoluminescence intensity $I_{pl}$ is larger when $m_S$=0 than when $m_S$=±1.

We study individual NV centers that were artificially created in a high-purity, single crystal diamond grown by chemical vapor deposition. The diamond was implanted with isotopically pure nitrogen-15 and then processed to form single NV centers [18]. These centers can be individually addressed via our confocal photoluminescence microscope that has a spatial resolution of ~400 nm [19**Error! Bookmark not defined.**]. Figure 1(b) shows a spatial $I_{pl}$ map of the region of the sample containing two of the NV centers we studied.

Spin rotations are induced by applying a microwave driving field $B_{rf}$ of up to 10 G, corresponding to Rabi frequencies of up to 14 MHz [20], which is large enough to cause a significant transition probability between different levels during the short (~8-12 ns) [16] ES lifetime (Fig. 1(a)).

Electron spin resonance (ESR) spectroscopy is performed by sweeping the frequency of $B_{rf}$ under continuous wave (CW) optical illumination and measuring $I_{pl}$. When the driving field is not resonant with a spin transition, the laser illumination polarizes the NV center into the $m_S$=0 spin state, resulting in a maximum value of $I_{pl}$. In contrast, when $B_{rf}$ is resonant with a transition between the $m_S=0$ and $m_S=\pm 1$ levels, the spin is rotated out of $m_S=0$ and $I_{pl}$ is reduced due to the lower photoluminescence rate in $m_S=\pm 1$. If the optical transition is spin-conserving, spin rotations in both the GS and the ES can be detected. The upper trace in Fig. 1 (c) shows a typical CW ESR measurement of NV1 taken at B=57 G. In addition to the two sharp resonances that correspond to the well-known GS transitions into $m_S=\pm 1$ from $m_S=0$, there are two additional, broad resonances that we attribute to spin transitions in the orbital excited state.

In order to verify that the broad features are associated with ES spin transitions, we repeat the measurement with $B_{rf}$ only applied when the NV center is in the GS. First, the NV spin is initialized into the $m_S=0$ state by optical pumping for 8 µs, after which the laser is turned off. Following a 3 µs wait time, which ensures that the NV center has relaxed to the GS, a 500 ns burst of $B_{rf}$ is applied. Finally, with $B_{rf}$ off, the laser and the photon counters are turned on for 1 µs to read-out the spin state. This sequence is repeated many (~$10^5$) times at each frequency to build up statistics. With this procedure, only spin transitions in the ground state are probed. The result is plotted in figure 1(d) below the previously described CW measurement. The broad resonances are not visible when the spin is only manipulated in the GS, which confirms that they correspond to spin transitions in the ES.

We now apply our technique to study the ES spin structure in more detail. Figure 1(d) shows a surface-relief plot of CW ESR data of NV1 taken as a function of frequency as well as magnetic field $B$ closely aligned along the [111] crystal axis of the sample [21]. These data have been inverted so that dips in $I_{pl}$ appear as peaks in order to make the features easier to follow. The familiar GS spin transitions between $m_S=0$ and $m_S=+1$ (-1) appear as a series of sharp peaks starting from the GS zero field spin-splitting of 2.87 GHz and tracking to higher (lower) frequency as $B$ increases. The ES resonances have a different zero-field splitting (about 1.43 GHz) and shift along a path essentially parallel to each of the GS lines, indicating that the g-factor of the ES, $g_{es}$, is similar to the value for the GS.

At $B$=500 G there is a high-amplitude and broad feature where the ES $m_S$=-1 spin level crosses the energy of the $m_S$=0 level. Previously, Epstein et al. [19] observed a characteristic decrease in $I_{pl}$ under CW optical illumination around 500 G. Our data explains this feature: the spin polarization (and thus $I_{pl}$) is reduced by mixing of the $m_S$=0 and $m_S$=-1 levels in the ES around the degeneracy at 500 G. This is analogous to the situation at 1028 G where $m_S$=0 and $m_S$=-1 cross in the GS [19].

Careful examination of the ES dips reveals that each resonance is actually composed of two broad dips (Fig. 2(a)-(c)). This is consistent with the hyperfine (HF) splitting of the ES spin levels by the nuclear spin of $^{15}$N (I=1/2). Similarly, the GS transition is also HF split into a doublet with a splitting of 3.01±0.05 MHz at 31 G (not shown), which is similar to previous reports for $^{15}$NV centers [22]. The HF splitting in the ES, however, is significantly larger (~60 MHz) which reflects the larger spin density of the ES wavefunction near the N nucleus [23, 24] and agrees with a rough theoretical estimate. Identification of HF split transitions allows us to label the ES spin levels according to the electronic and nuclear spin quantum numbers of the NV electronic spin ($m_S$) and the $^{15}$N nuclear spin ($m_I$), respectively. At $B$>50 G and away from the 500 G ES level crossing, these are good quantum numbers and we will use them to label specific transitions.

In order to quantify the evolution of these transitions as B is increased, we fit high-resolution scans of the ES dips with two Lorentizians. Figure 2(a)-(c) show example fits at different fields while figures 2(d)-(e) contain some of the fit results for the $m_S$=0↔+1transitions. Figure 2(d) shows the amplitudes of the $m_I$=-1/2 dips (higher frequency) [25] relative to the $m_I$=+1/2 dips (lower frequency). These data show a large nuclear polarization, possibly caused by dynamic nuclear polarization through the hyperfine interaction. This will be examined in more detail in future work. A nuclear polarization of the opposite sign has also been observed at 490 G in GS transitions in Ref. [1].

The large polarization causes many of the $m_I=-1/2$ dips to be partially obscured by the $m_I=+1/2$ dips (e.g. see Fig. 2(b)) which makes determination of the FWHM of the $m_I=-1/2$ transitions unreliable for a range of fields. Hence we plot only the widths of the $m_I=+1/2$ transitions in Fig. 2(e), which shows a value around 80 MHz. Qualitatively similar features are present in the $m_S=0\leftrightarrow-1$, $m_I=\pm1/2$ features (not shown).

Figure 2 (f)-(h) illustrate how the ES dips vary with different experimental parameters. Figure 2(f) shows that the average ES dip amplitude increases with laser power. The amplitudes of the dips also increase linearly as the amplitude of $B_{rf}$ is ramped (Fig. 2 (g)) which confirms that the ES transitions are not yet saturated at these driving fields. Figure 2 (h) shows how the widths of the transitions change with increasing $B_{rf}$. The data indicate little dependence of the FWHM on $B_{rf}$, which suggests that the width is primarily determined by the lifetime of the states through ($\pi$ FWHM) $=(T_0)^{-1}+(T_{\pm1})^{-1} + (T^*)^{-1}$ where $T_0$=12 ns is the ES lifetime of $m_S$=0, $T_{\pm1}$=7.8 ns is the ES lifetime of $m_S=\pm1$ [16], and $T^*$ is a time associated with all other sources of decoherence and broadening by the driving field. Neglecting $T^*$, we expect a FWHM of ~67 MHz, which is in good agreement with the measured values (Fig. 2(d) and (f)) and further supports that the optical lifetimes are the dominant factor that determine the FWHM in these data.

The frequencies that result from the Lorentzian fits for each of the four ES dips are plotted together in figure 3(a) and (b). We simultaneously fit these data to four transitions between the eigenvalues of the ES spin Hamiltonian:

$$H = D_{es}S_Z^2 + \mu_B g_{es}\vec{B}\cdot\vec{S} + E_{es}(S_X^2 - S_Y^2) - A_{es}\vec{S}\cdot\vec{I} \qquad (1)$$

where $\vec{S}$ is the NV center spin-1 in the ES with $S_X$, $S_Y$, and $S_Z$ components and $\vec{I}$ is the nuclear spin-1/2 of $^{15}$N. As mentioned previously, $D_{es}$ is the ES longitudinal (zero-field) splitting [26], $g_{es}$ is the ES g-factor, $A_{es}$ is the hyperfine splitting constant, and $E_{es}$ is the coefficient of strain-induced transverse anisotropy. We expect the latter to be noticeable: the ES, as an orbital doublet, is strongly affected by strain. Moreover, at $B$=0.7 G we also see that the orbital singlet in the GS is strain-split by ~6 MHz more than can be accounted for by Zeeman splitting (Fig. 3(d) inset). Since the GS is only weakly affected by strain, the strain on this NV center is large. Fitting the data we find $D_{es}$=1.425±0.003 GHz, $g_{es}$=2.01±0.01, $A_{es}$=61±6 MHz and $E_{es}$=70±30 MHz. The value of $g_{es}$ is, within our margin of error, identical to $g_{gs}$, which means that the orbital contribution, previously estimated to be ~0.1 [27, 28], is quenched in the case of high strain as expected. The best fits to the data are also shown in Fig. 3(a)-(b) as dashed lines.

Figure 3 (c) shows a schematic diagram of the excited state levels. The left diagram shows the role of strain in the *x*-direction at $B$=0 in presence of spin-orbit and spin-spin coupling (*z*-axis strain is not considered: it shifts the energy of all levels by the same amount). At large strain, two orbital branches are formed that are separated by tens of GHz and have $E_X$ and $E_Y$ symmetry. The spin-orbit/spin-spin couplings split the spin levels inside each branch into levels of $S_Z$ ($m_S$=0), $S_X$, and $S_Y$ symmetry. When a magnetic field is applied (right diagram), the $S_X$, and $S_Y$ levels are mixed, forming the states well described with $m_S$=±1 quantum numbers due to Zeeman splitting (HF coupling omitted). At large strain and large magnetic field we would expect four HF-split doublets (two from each branch), while we observe only two. In Fig. 3(c), we tentatively identify the observed ES features as belonging to the upper orbital branch of the strain-split levels. This assignment is based on the energy scale of the observed features. The symmetry of the data, the agreement with Eq (1), and the value of the splitting within each of the doublets strongly supports our conclusion that the doublets are the result of HF splitting rather than belonging to separate branches of the diagram in figure 3(c). There are many factors that may explain why we do not see evidence of the lower branch. One possibility is spin mixing of $m_S$=±1 with $m_S=0$ levels due to transverse spin-orbit coupling [10]. If the ESR-induced transition probability into $m_S=\pm1$ is comparable to the spin mixing, dips in $I_{pl}$ would be undetectable. Other possibilities include a very short (<2 ns) lifetime, an unequal excitation of the orbital branches or a combination of these.

Finally, as further evidence that local strain is playing a dramatic role in the ES spin levels, we have plotted in figure 3(d) a CW ESR measurement at 0.7 G for another NV center, NV2, below the same measurement of NV1. Although NV2 has a value of $D_{es}$=1.425±0.001 GHz, nominally the same as NV1, the 0.7 G splitting between the ES dips of NV2 (552±1 MHz) is significantly larger than the splitting observed in NV1 (131±3 MHz), which may correspond to a larger strain-induced transverse anisotropy [29]. The low-field splitting between the GS spin levels in NV2 is also correspondingly larger than in NV1 (Fig. 3(d) insets).

In conclusion, we have presented a measurement of the excited state spin levels of a single NV center. We find that the data fit closely to a spin Hamiltonian including the effects of hyperfine splitting and transverse anisotropy caused by strain. We also present evidence that suggests that strain plays a critical role in the ES spin Hamiltonian which presents an opportunity to adjust the spin levels through strain engineering. Moreover, an electric field can also act as a local strain [15] which opens the door to high-speed manipulation of the ES spin-levels and may prove useful in the implementation of quantum gates as well as ultra-fast, all-optical quantum control schemes [13].

We thank Elias Sideras-Hiddad for supplying the diamond substrate as well as for conversations regarding implantation. This work is supported by AFOSR (G.D.F., D.D.A.), FOM and NWO (R.H.) and the Director, Office of Science, of the US DOE under Contract No. DE-AC02-05CH11231 (A.B., C.D.W., T.S.). Work at the Ames Laboratory was supported by the US DOE – Basic Energy Sciences under contract No. DE-AC02-07CH11358 (V.V.D.).

[1] T. Gaebel, *et al.*, Nat. Phys. **2**, 408 (2006).

[2] F. Jelezko, *et al.*, Appl. Phys. Lett. **81**, 2160 (2002).

[3] J. Harrison, M. J. Sellars, and N. B. Manson, J. Lumin. **107**, 245 (2004).

[4] A. Gruber, *et al.*, Science **276**, 2012 (1997).

[5] R. Hanson, *et al.*, Phys. Rev. Lett **97**, 087601 (2006).

[6] R. Hanson, *et al.*, Science **320**, 352 (2008).

[7] F. Jelezko, *et al.*, Phys. Rev. Lett **93**, 130501 (2004).

[8] L. Childress, *et al.*, Science **314**, 281 (2006).

[9] M. V. G. Dutt, *et al.*, Science **316**, 1312 (2007).

[10] N. B. Manson, J. P. Harrison and M. J. Sellars, Phys. Rev. B **74**, 104303 (2006).

[11] J. A. Larrson and P. Delaney, Phys. Rev. B. **77**, 165201 (2008).

[12] N. B. Manson and R. L. McMurtrie, J. Lumin. **127**, 98 (2007).

[13] C. Santori, *et al.*, Phys. Rev. Lett. **97**, 247401 (2006).

[14] L. Childress, *et al.*, Phys. Rev. Lett. **96**, 070504 (2006).

[15] P. Tamarat, *et al.*, N. J. Phys. **10**, 045004 (2008).

[16] A. Batalov, *et al.*, Phys. Rev. Lett. **100**, 0077401 (2008).

[17] J. Loubser and J. A. van Wyk, Rep. Prog. Phys. **41**, 1201 (1978).

[18] T. Schenkel, *et al.*, in preparation.

[19] R. J. Epstein, *et al.*, Nat. Phys. **1**, 94 (2005).

[20] We use $B_{rf}$~7 G between 0.01-0.8 GHz due to a component power limitation.

[21] We note that the sample is only exactly aligned with [111] at high field, and there is a small (up to 3°) misalignment at lower fields. Although the misalignment is taken into account in all calculations and fits, it has a negligible effect on the results and conclusions of this investigation.

[22] J. R. Rabeau, *et al.*, Appl. Phys. Lett. **88**, 023113 (2006).

[23] A. Lenef and S. C. Rand, Phys. Rev. B **53**, 13441 (1996).

[24] A. Gali, M. Fyta, and E. Kaxiras, Phys. Rev. B **77**, 155206 (2008).

[25] The sign of the contact hyperfine Hamiltonian is negative due to the negative gyromagnetic ratio of the $^{15}$N nuclear spin (C. P. Slicter *Principles of Magnetic Resonance* (Springer, New York, 1990) 3$^{rd}$ ed). Therefore, ($m_S$, $m_I$) with the same sign has lower energy than opposite sign.

[26] We have assumed that the ES $S_X$ and $S_Y$ levels have energy above $S_Z$ as in the unstrained case. This is in agreement with ref. [10] and [15].

[27] J. P. D. Martin, J. Lumin. **81**, 237 (1999).

[28] N. R. S. Reddy, E. R. Krausz, and N. B. Manson, J. Lumin. **38**, 46 (1987).

[29] We note that NV2 is aligned with a different <111> symmetry direction than NV1.

Figure Captions

Figure 1. (a) Energy diagram of electronic and spin levels for a NV center. Both the ground state ($^3A$) and the excited state ($^3E$) are spin triplets, with a zero field splitting ($D_{gs}$, $D_{es}$) between the $m_S=0$ and the $m_S=\pm 1$ levels. Black arrows show the largely spin-conserving optical transitions. There is also non-radiative relaxation through $^1A$ that is not spin conserving (Gray arrows indicate the dominate pathway) (b) Spatial photoluminescence map with NV1 and NV2 indicated. We verified that each spot corresponds to a single NV center using photon correlation measurements (not shown). (c) The top trace shows optically detected spin resonance at $B=57$ G where swept frequency $B_{rf}$, laser illumination, and photon counters are all on continuously. The bottom trace is a measurement where $B_{rf}$ is applied while the laser is off so that the spin is manipulated only in the ground state. (d) Surface plot of CW spin resonance as a function of magnetic field. The viewpoint is below the data so that dips in $I_{pl}$ appear as peaks. The sharp features are the GS spin resonances while the broad features are resonant transitions in the ES.

Figure 2. Fits of the ES $m_S=0\leftrightarrow+1$ doublets to a double Lorentzian for (a) 67 G, (b) 503 G and (c) 804 G. (d) FWHM vs. $B$ for $m_S=0\leftrightarrow+1$, $m_I=+1/2$. (e) Amplitudes of the $m_S=0\leftrightarrow+1$ dips vs. $B$. (f) Average variation of dip amplitude vs. laser power. (g)-(h) Fit results for $m_S=0\leftrightarrow-1$, $m_I=-1/2$ (●), $m_S=0\leftrightarrow-1$, $m_I=+1/2$ (◆), $m_S=0\leftrightarrow+1$, $m_I=+1/2$ (▲) and $m_S=0\leftrightarrow+1$, $m_I=-1/2$ (■) at $B=57$ G. (g) Amplitude of the dips in $I_{pl}$ vs. amplitude of $B_{rf}$. (h) FWHM in MHz vs. amplitude of $B_{rf}$.

Figure 3. Frequencies of the ES spin transitions obtained from Lorentzian fits to the spin resonance data (points) as well as fits to the eigenvalues of to Eq. 1 (dashed lines) as described in the text for the entire field range (a) and low field (b). (c) Left: diagram of the energy levels of the excited $^3E$ maniforld as a function of strain along *x*-axis. Adopting the notation of [10]: $A_1=|E_X, S_X\rangle+|E_Y, S_Y\rangle$, $A_2=|E_X, S_Y\rangle-|E_Y, S_X\rangle$, $E'_Y=|E_X, S_Z\rangle$, $E'_X=|E_Y, S_Z\rangle$, $E_Y=-|E_X, S_X\rangle+|E_Y, S_Y\rangle$, and $E_X=-|E_X, S_Y\rangle-|E_Y, S_X\rangle$. At large strain this orbital doublet is split into two orbital singlet branches ($E_X$ and $E_Y$ symmetry), each comprised of three spin levels. Right: diagram of level splitting as a function of *B*, assuming the spin features are associated with $E_X$. (d) CW ESR data at B=0.7 G for NV1 and NV2. The insets show the splitting of the GS transitions at 0.7 G by ~ 9 MHz and 21 MHz for NV1 and NV2 respectively (marked with arrows). Since only ~2 MHz of the splitting can be due to the Zeeman effect in a 0.7 G field, this indicates there is strain-induced splitting between $S_X$ and $S_Y$ even in the GS orbital singlet. For NV1, we used a small optical power and $B_{rf}$ to also resolve the hyperfine splitting in the GS transitions. For NV2, however, a larger $B_{rf}$ was used due to the large asymmetry between the $S_Z\leftrightarrow S_X$ and $S_Z\leftrightarrow S_Y$ transitions.

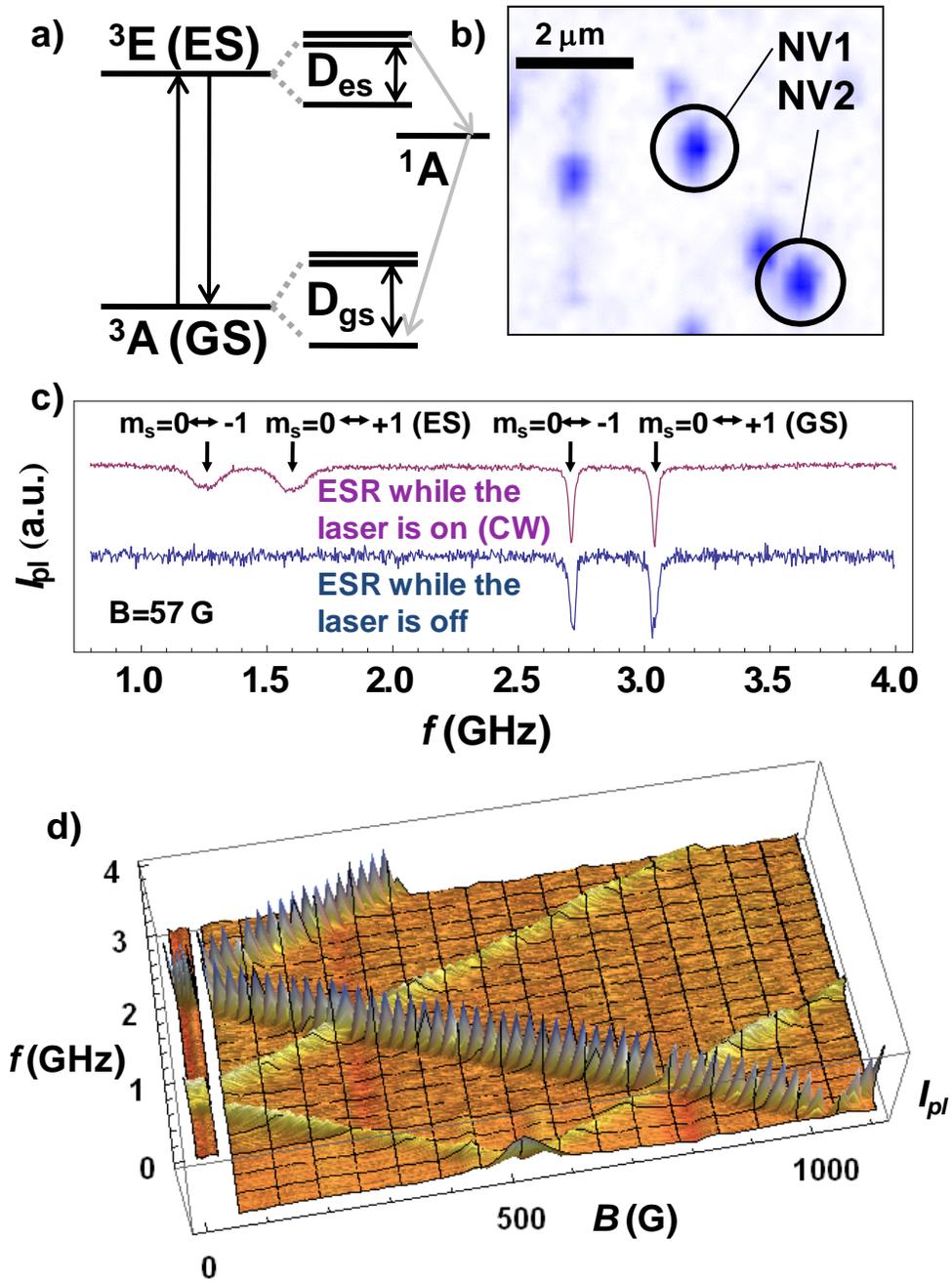

Figure 1

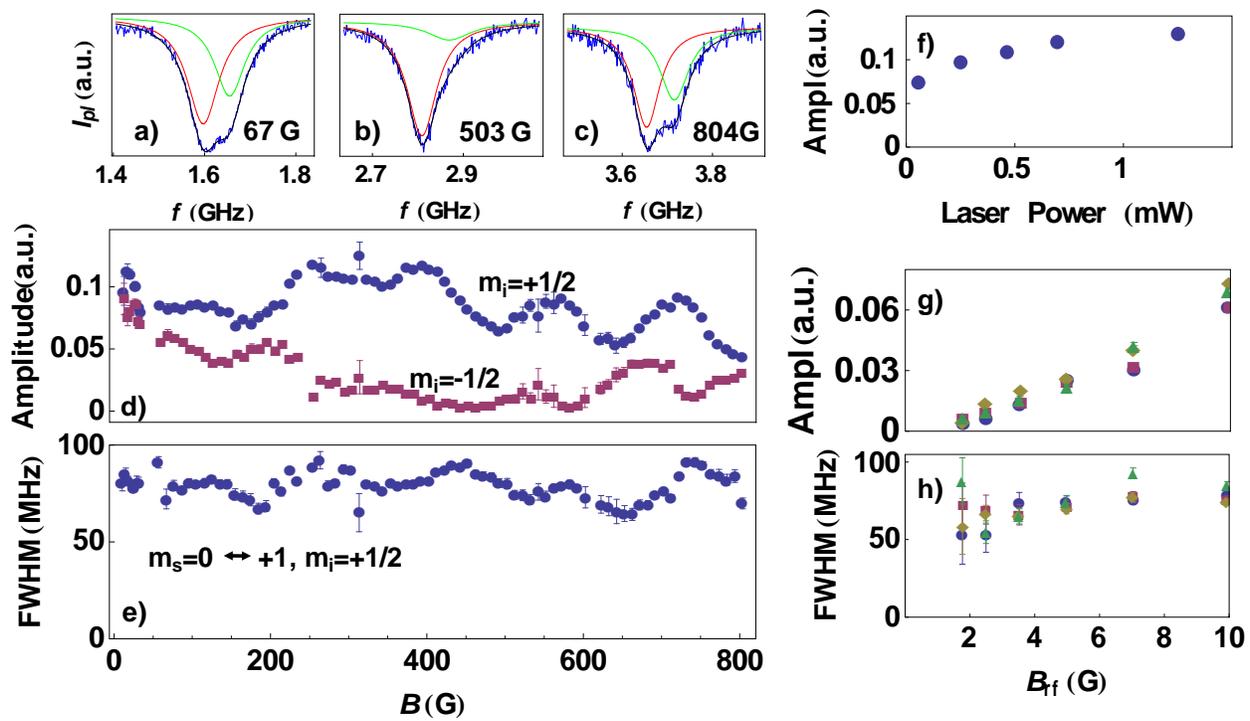

Figure 2

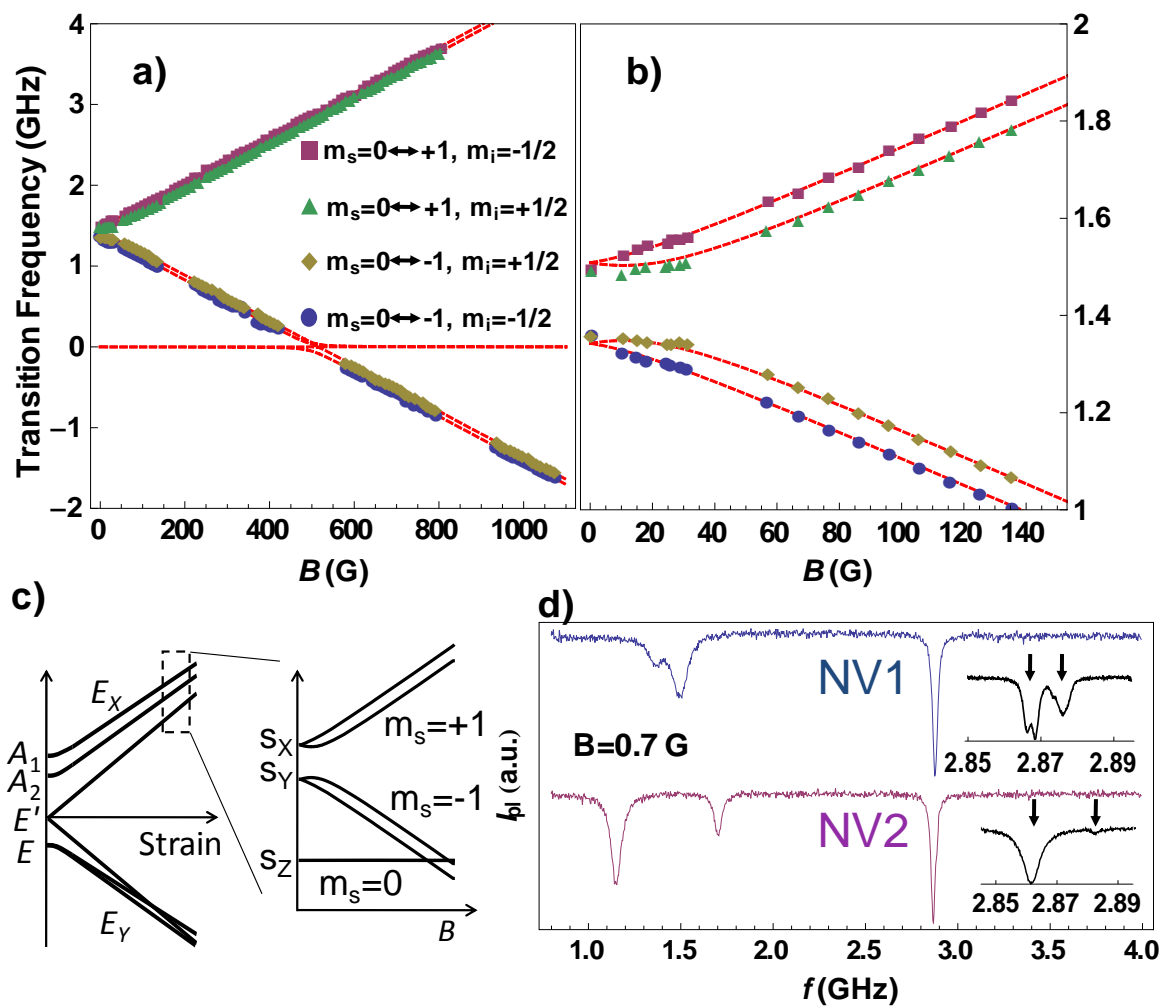

Figure 3